\documentstyle[12pt]{article}
\renewcommand{\theequation}{{\thesection.\arabic{equation}}}
\newcommand{\ol}[1]{\overline{#1}}

\makeatletter
\def\eqnarray{%
  \stepcounter{equation}%
  \let\@currentlabel=\theequation
  \global\@eqnswtrue
  \global\@eqcnt\z@
  \tabskip\@centering
  \let\\=\@eqncr
  $$\halign to \displaywidth\bgroup\@eqnsel\hskip\@centering
  $\displaystyle\tabskip\z@{##}$&\global\@eqcnt\@ne
  \hfil$\displaystyle{{}##{}}$\hfil
  &\global\@eqcnt\tw@$\displaystyle\tabskip\z@{##}$\hfil
  \tabskip\@centering&\llap{##}\tabskip\z@\cr}
\makeatother

\title{
Bosonization of Thirring Model
\\ in Arbitrary Dimension
}
\author{%
  {\sc Kenji Ikegami},
  {\sc Kei-ichi Kondo}
and {\sc Atsushi Nakamura}
  \thanks{e-mail: ikegami, kondo,
nakamura@cuphd.nd.chiba-u.ac.jp}
\\  \vspace{0.5cm} \\
  {\it Department of Physics,}\\
  {\it Faculty of Science,} \\
  {\it Chiba University, Chiba 263, Japan}
  \vspace{1.2em} \\
  }
\date{CHIBA-EP-90 \\
      June 1995 \\
      hep-th/9509173}
\begin{document}
\maketitle
\centerline{Abstract}
We propose to use a novel master Lagrangian for performing
the bosonization of the $D$-dimensional massive Thirring
model in $D=d+1 \ge 2$ dimensions.  It is shown that our
master Lagrangian is able to relate the previous
interpolating Lagrangians each other which have been
recently used to show the equivalence of the massive
Thirring model in (2+1) dimensions with the
Maxwell-Chern-Simons theory.  Starting from the phase-space
path integral representation of the master Lagrangian, we
give an alternative proof for  this equivalence up to the
next-to-leading order in the expansion of the inverse
fermion mass.  Moreover, in (3+1)-dimensional case, the
bosonized theory is shown to be equivalent to the massive
antisymmetric tensor gauge theory. As a byproduct, we
reproduce the well-known result on bosonization of the
(1+1)-dimensional Thirring model following the same
strategy. Finally a possibility of extending our strategy to
the non-Abelian case is also discussed.

\newpage

\section{Introduction}
\setcounter{equation}{0}
\par
In this paper we investigate the bosonization
of the
Thirring model \cite{Thirring} as a gauge theory
\cite{IKSY95,Kondo95th1,Kondo95th2} in $D=d+1$ spacetime
dimensions  ($D \ge 2$). As is well known, a lot of works
have been devoted to the bosonization of the
(1+1)-dimensional fermionic model, see
\cite{Coleman,AA95,BQ94,Kondo95th2} and references therein for the
Thirring model.   However the bosonization is not necessarily
restricted to the (1+1) dimensional case.  Actually the
fermion-boson equivalence was discussed earlier e.g. in
\cite{DR88}. Moreover the bosonization of fermion systems
in $D > 2$ dimensions has regained interest by recent works
\cite{BLQ94,FS94,IKSY95,Kondo95th2}.
Particularly the bosonization recipe for abelian systems in
$D=3$ was devised in \cite{BLQ94,FS94,IKSY95,Kondo95th2}.
We start from one of the reformulations of the Thirring
model as a gauge theory which is first proposed by Itoh et
al. \cite{IKSY95} and subsequently one of the
authors (K.-I. K) \cite{Kondo95th1,Kondo95th2} from
a different viewpoint. The basic idea of introducing the
gauge degrees of freedom in the bosonization was earlier
proposed in more general form in
\cite{BQ94,BLQ94}, although the works
\cite{IKSY95,Kondo95th1,Kondo95th2} were done
independently. In the vanishing coupling limit $G
\rightarrow 0$, our reformulation of the Thirring model
reduces to the result of \cite{BQ94,BLQ94}.
\par
\par
In this paper we consider the Thirring model defined by
the Lagrangian:
\begin{eqnarray}
 {\cal L}_{Th}
 = \bar \psi^a i \gamma^\mu \partial_\mu \psi^a
 - m_a \bar \psi^a \psi^a
- {G \over 2N}
(\bar\psi^a \gamma_\mu \psi^a)
(\bar\psi^b \gamma^\mu \psi^b),
\label{th}
\end{eqnarray}
where $\psi^a$ is a Dirac spinor and the indices
$a, b$ are summed over from $1$ to $N$, and
$\gamma_\mu (\mu = 0,..., D-1)$  are gamma matrices
satisfying the Clifford algebra,
$\{ \gamma_\mu, \gamma_\nu \} = 2 g_{\mu\nu} {\bf 1}
= 2 {\rm diag}(1,-1,...,-1)$.
As usual, by introducing an auxiliary vector field
$A_\mu$, the Thirring model is equivalently rewritten as
\begin{eqnarray}
 {\cal L}_{Th'}
 = \bar \psi^a i \gamma^\mu
 D_{\mu}[A]
 \psi^a - m_a \bar \psi^a \psi^a
 + {1 \over 2G} A_\mu A^\mu .
 \label{th'}
\end{eqnarray}
where
\begin{equation}
D_{\mu}[A]\equiv
\partial_\mu - {i \over \sqrt{N}} A_\mu .
\end{equation}
\par
The fermionic degrees of freedom can be integrated away
from the Lagrangian (\ref{th'}). Especially, in the massive
fermion case, the fermion determinant leads to a local
expression for ${\cal L}_G[A]$ of the field $A_\mu$
defined by
\begin{eqnarray}
 \ln \det  \left( {i \gamma^\mu D_{\mu}[A] + m_a \over
 i \gamma^\mu \partial_\mu + m_a}  \right)
 =  \int d^3 x {\cal L}_G[A]
 + {\cal O} \left( {\partial^2 \over |m_a|^2} \right).
 \label{det}
\end{eqnarray}
The original Thirring model (\ref{th}) has no gauge
symmetry and this is the case even after the auxiliary field
is introduced in (\ref{th'}). However, if we are
allowed to identify the vector field $A_\mu$ with the gauge
field and, at the same time, able to adopt an appropriate
gauge-invariant regularization scheme, the resulting ${\cal
L}_G[A]$ in (\ref{det}) leads to the gauge invariant
expression.  For example, in 2+1 dimensions, we have
\begin{eqnarray}
 {\cal L}_G[A] =
 {i \theta_{CS} \over 4}
 \epsilon^{\mu\nu\rho} A_\rho F_{\mu\nu}
 - {1 \over 24\pi |m|} F_{\mu\nu} F^{\mu\nu} ,\label{eq:LG}
\end{eqnarray}
with
\begin{eqnarray}
 \theta_{CS} = {1 \over N} \sum_{a=1}^{N} sgn(m_a)
 {1 \over 4\pi},
\end{eqnarray}
where $|m_a|=m$ for all $a$ is assumed \cite{Kondo95th2}.
For this scenario to be successful, the Lagrangian
\begin{equation}
{\cal L}_G[A] + {1 \over 2G} A_\mu^2,\label{selfdual}
\end{equation}
which is self-dual model, for $A_\mu$ must be gauge
invariant. This is realized by identifying the Thirring model
as a gauge-fixed version of some gauge theory
\cite{Kondo95th1} by following the Batalin-Fradkin method
\cite{BF} based on the general formalism of
Batalin-Fradkin-Vilkovisky (BFV)
\cite{BFV} for constraint system.  Here the requirement of
gauge invariance plays the role of selecting a class of
gauge-invariant regularizations and of removing some
ambiguities related to the regularization
\cite{Kondo95th1}.

\par
Keeping the above remarks in mind, we briefly review the
recent development on bosonization.  It has been shown that
the (2+1)-dimensional massive Thirring model is equivalent
to the Maxwell-Chern-Simons (MCS) theory, up to the leading
order \cite{FS94} and to the next-to-leading order in
$1/|m|$ \cite{Kondo95th2}.
\footnote{Note that the bosonization of free fermion model
in (2+1) dimensions reduces to the Chern-Simons theory
(without the Maxwell term) \cite{BLQ94}.  In our formalism,
this case is reproduced as the free fermion limit, $G
\rightarrow 0$.}   This fact was first shown by way of the
interpolating Lagrangian \cite{FS94}.  However the
interpolating Lagrangians adopted by two papers
\cite{FS94,Kondo95th2} are different from each other.
Fradkin and Schaposnik \cite{FS94} uses the interpolating
Lagrangian of the form:
\begin{eqnarray}
 {\cal L}_{FS}[V,H]
 =  {1 \over 2G} V^\mu V_\mu
 - {1 \over 2}\epsilon^{\mu\nu\rho} V_\rho F_{\mu\nu}[H]
 + 2\pi \epsilon^{\mu\nu\rho} H_\mu \partial_\nu
H_\rho ,
\label{interFS}
\end{eqnarray}
where $F_{\mu\nu}[H]$ is the field strength for the gauge
field $H_\mu$.
They start from the observation that the Lagrangian of
the Thirring model written in terms of the auxiliary
field $V_\mu$ (corresponding to $A_\mu$ in eq.~(\ref{det}))
up to the leading order of
$1/m$ is equal to the self-dual Lagrangian introduced by
Townsend, Pilch and van Nieuwenhuizen
\cite{TPN84} and Deser and Jackiw \cite{DJ84}:
\begin{eqnarray}
 {\cal L}_{SD}[V] =  {1 \over 2G}  V_{\mu} V^{\mu}
+ {\cal L}_G[V].
\label{SD}
\end{eqnarray}
Then they find the master Lagrangian (\ref{interFS}) such
that the self-dual Lagrangian (\ref{SD}) is obtained by
integrating out the $H_\mu$ field.
The MCS theory with the Lagrangian
\begin{equation}
{\cal L}_{MCS}[H]=
   -\frac{G}{4}F_{\mu\nu}[H]F^{\mu\nu}[H]
   -2\pi \epsilon^{\mu\nu\rho}H_{\mu}\partial_{\nu}H_{\rho},
   \label{MCS}
\end{equation}
is obtained by integrating away the
field $V_\mu$ from ${\cal L}_{FS}[V,H]$.
It should be noted that the interpolating Lagrangian
(\ref{interFS}) is invariant under the transformation:
$
\delta H_\mu = \partial_\mu \omega,
$
$
\delta V_\mu = 0 .
$
 However the field $V_\mu$ does not have the gauge
invariance, although they use the gauge invariant
expression (\ref{det}) for ${\cal L}_G[V]$ as if the
field $V_\mu$ was the gauge field.
\par
On the other hand, Kondo's strategy \cite{Kondo95th2} is in
sharp contrast to  the treatment of Fradkin and Schaposnik
\cite{FS94}.  He adopts the interpolating Lagrangian:
\begin{eqnarray}
 {\cal L}_{K}[A,H] =   - {G \over 4}
 F_{\mu\nu}[H]  F^{\mu\nu}[H]
+ {1 \over 2} \epsilon^{\mu \nu \rho}
   F_{\mu\nu}[H]  A_{\rho}
+ {\cal L}_G[A] ,
\label{interK}
\end{eqnarray}
which was {\it derived} by starting from the reformulation
of the Thirring model as a gauge theory.
As a result, the interpolating Lagrangian
${\cal L}_{K}[A,H]$ is invariant under two
independent gauge transformations:
$
\delta A_\mu = \partial_\mu \lambda,
$
and
$
\delta H_\mu = \partial_\mu \omega .
$
\footnote{
Quite recently the interpolating Lagrangian (\ref{interK})
was used to show the equivalence to all orders in the
inverse fermion mass by Banerjee \cite{Banerjee95}.
}
However the connection of two approaches was not
necessarily clear at that stage.
\par
In this paper we propose to use the following master
Lagrangian in order to investigate the bosonization of the
$D$-dimensional massive Thirring model ($D \ge 2$):
\begin{equation}
  {\cal L}_{M}[A,H,K]
  =  {1 \over 2G}(A_\mu-K_\mu)^2
  + {1 \over 2}\epsilon^{\mu_1 ... \mu_D} H_{\mu_3 ...
\mu_D} F_{\mu_1\mu_2}[K]
+ {\cal L}_G[A],
\label{master1}
\end{equation}
where $H_{\mu_3 ... \mu_D}$ is anti-symmetric tensor field
of rank $D-2$ for $D> 3$, vector field $H_{\mu}$ for $D=3$ and
a scalar field
$H$ for
$D=2$. After redefinition of the field variable, the master
Lagrangian~(\ref{master1}) has another form:
\begin{equation}
 {\cal L}_M'[A,H,V] =
 {1 \over 2G}(V_\mu)^2
 + {1 \over 2} \epsilon^{\mu_{1}...\mu_{D}}
H_{\mu_{3}...\mu_{D}} F_{\mu_{1}\mu_{2}}[V+A]
  + {\cal L}_G[A] .
\label{master2}
\end{equation}
\par
An advantage of the master Lagrangian (\ref{master1}) or
(\ref{master2}) is that it is able to interpolate two types
of apparently different interpolating Lagrangians: for
example, (\ref{interFS}) and (\ref{interK}) in (2+1)
dimensions.  Actually, we show from the master Lagrangian
(\ref{master2}) for $D=3$ that
${\cal L}_{FS}[V,H]$ is obtained by integrating out
the $A_\mu$ field, while ${\cal L}_{K}[A,H]$ is
obtained by integrating out the $V_\mu$ field.
On the other hand, after eliminating the field
$H_\mu$ and $V_\mu$, we get (see Figure 1):
\begin{eqnarray}
 {\cal L}_{Th''}[A,\theta] =
 {1 \over 2G}(A_\mu - \sqrt{N} \partial_\mu \theta)^2
 + {\cal L}_G[A].\label{Th''}
\end{eqnarray}
This is nothing but a gauge-invariant formulation of the
Thirring model with the field $\theta$
being identified with the St\"uckelberg field.
This Lagrangian was the starting point of the
gauge-invariant formulation.  The Lagrangian (\ref{Th''})
should be compared with the self-dual Lagrangian
(\ref{SD}). This difference comes from the fact that our
master Lagrangian (\ref{master2}) has independent gauge
invariance for two {\it gauge} fields
$A_\mu$ and
$H_\mu$, while this is not the case for $V_\mu$:
\begin{eqnarray}
\delta A_\mu = \partial_\mu \lambda,
\quad
\delta H_\mu = \partial_\mu \omega ,
\quad
\delta V_\mu = 0.
\label{symm}
\end{eqnarray}
\par
The classical equivalence of the master
Lagrangian with the non-linear $\sigma$-model is
easy to understand.  Indeed the master Lagrangian
(\ref{master2}) is a polynomial formulation of the
{\it gauged} non-linear $\sigma$-model
\cite{Townsend79,FT81,TM90}:
\begin{eqnarray}
 {\cal L}_{gNL\sigma}[\varphi, A]
 =   (D_\mu[A] \varphi)^\dagger (D^\mu[A] \varphi)
 + {\cal L}_G[A] ,\label{NLSM}
\end{eqnarray}
with a local constraint:
$\varphi(x) \varphi^*(x) = {N \over 2G}$.
\footnote{
As a special case, putting $A_\mu=0$ in the master
Lagrangian, we obtain the polynomial formulation
\begin{eqnarray}
  {\cal L}_{P}[H,K]
  =  {1 \over 2G}(K_\mu)^2
  + {1 \over 2}\epsilon^{\mu_1 ... \mu_D} H_{\mu_3 ...
\mu_D} F_{\mu_1\mu_2}[K],
\label{master3}
\end{eqnarray}
of the non-linear $\sigma$-model
\cite{DFT92,FT94,FM95} with the Lagrangian
\begin{eqnarray}
{\cal L}_{NL\sigma}[\varphi]
= (\partial_\mu \varphi)^\dagger (\partial^\mu \varphi) .
\end{eqnarray}
}
\par
Nevertheless, it is not necessarily straightforward to show
the quantum equivalence.  In particular, we must be rather
careful in treating the non-Abelian case, which will be
discussed in the final section. The bosonization
\cite{BLQ94,FS94,IKSY95,Kondo95th2} in $D > 2$
dimensions has been carried out based on the configuration
space path-integral expressions of the partition
functions.   In this paper we show the equivalence between
the (2+1)-dimensional massive Thirring model and the
Maxwell-Chern-Simons theory by starting from the
phase-space path integral representation of our master
Lagrangian.
This equivalence is shown up to the leading order of $1/m$
in section 2.1 and up to the next-to-leading in section
2.2. This type of investigation is very important to
elucidate the constraint structure of the various
Lagrangian in question.  Such a strategy was taken  in
\cite{BRR95} for the master Lagrangian of Deser and Jackiw
\cite{DJ84} to study the connection between the self-dual
model and the Maxwell-Chern-Simons theory,
which is now included as a part of our investigation in the
following.
In section 3, our method is applied to the (1+1)
dimensional case and we reproduce the previous result on
the bosonization of (1+1)-dimensional Thirring model.
The case of $D \ge 4$ is discussed in section 4.
The final section is devoted to conclusion and discussion.

%%%%%%%%%%%%%%%%% Main Part
\section{(2+1) dimensions}
\setcounter{equation}{0}
\subsection{up to the leading order}
\par
In order to demonstrate the relation among the
massive Thirring model, the self-dual model, the
Maxwell-Chern-Simons theory and the non-linear
$\sigma$ model,  we take into account
${\cal L}_G(A)$ defined in (\ref{eq:LG}) up to the
leading order of $\frac{1}{m}$:
\begin{equation}
{\cal L}_G[A]
    = \frac {i\theta_{CS}}{2}
   \epsilon^{\mu\nu\rho}A_{\mu}\partial_{\nu} A_{\rho}.
\end{equation}
Hence the master Lagrangian (\ref{master2}) up to the
leading order reads
\begin{equation}
{\cal L}_{3Leading}[A,H,V]  =
   \frac {i\theta_{CS}}{2}
   \epsilon^{\mu\nu\rho}A_{\mu}\partial_{\nu} A_{\rho}
     +  \frac{1}{2G}  V_{\mu}V^{\mu}
     +\frac{1}{2}\epsilon^{\mu\nu\rho}
        H_{\mu}F_{\nu\rho}[A+V].
     \label{3L}
\end{equation}
The Lagrangian (\ref{3L}) has the primary constraints:
\begin{eqnarray}
\phi^0_A &\equiv& \pi^0_A \approx  0,\quad
  \phi^i_A \equiv\pi^i_A
    -\epsilon^{ij}(\frac {i\theta_{CS}}{2} A_j+H_j)\approx 0,
    \label{piA}
               \cr
\phi^0_V &\equiv& \pi^0_V \approx  0,\quad
  \phi^i_V \equiv   \pi^i_V -\epsilon^{ij}H_j\approx0,
               \cr
\phi^{\mu}_H &\equiv &\pi^{\mu}_H \approx 0,
\end{eqnarray}
where $\epsilon^{ij}\equiv \epsilon^{0ij}$. Then we obtain
the canonical Hamiltonian:
\begin{eqnarray}
{\cal H}_{3Leading} &=&
       -i\theta_{CS} \epsilon^{ij}A_0 \partial_i A_j
          -\frac{1}{2G}V^{\mu}V_{\mu}
          \cr
          &&
          -\frac{1}{2}\epsilon^{ij}H_0 F_{ij}[A + V]
          -\epsilon^{ij}H_i \partial_j (A_0 + V_0).
\end{eqnarray}
 Using this Hamiltonian, we get the secondary constraints:
\begin{eqnarray}
\Phi_A &\equiv&
  \epsilon^{ij}\partial_i(i\theta_{CS}A_j+H_j)\approx
0,\cr
\Phi_V &\equiv&
   \frac{1}{G}V_0 +
     \epsilon^{ij}\partial_i H_j\approx 0,\cr
\Phi_H &\equiv&
   \frac{1}{2}\epsilon^{ij}F_{ij}[A+V]\approx 0.
\end{eqnarray}
By redefining the constraints:
\begin{eqnarray}
\Phi_A'&\equiv& \Phi_A+\partial_i\phi^i_A,
\cr
\Phi_H'&\equiv& \Phi_H + \partial_i\phi^i_H,
\cr
\Phi'_V&\equiv& \Phi_V + \partial_i\phi^i_V,
\end{eqnarray}
the Poisson brackets are simplified so that the
resulting Poisson brackets of $\Phi_A', \Phi_H'$ with other
constraints vanish:
\begin{equation}
\{ \Phi'_A, \cdot\}=\{ \Phi'_H, \cdot\}=0.
\end{equation}
Then we enumerate the non-vanishing Poisson brackets:
\begin{eqnarray}
\begin{array}{@{\,}ll@{\,}}
\{ \Phi'_V(\vec{x},t), \phi^0_V(\vec{y},t)
\}=
\frac{1}{G}\delta^2(\vec{x}-\vec{y}),
&\nonumber
\\
\{ \phi^i_A(\vec{x},t), \phi^j_A(\vec{y},t) \}
      =
-i\theta_{CS}\epsilon^{ij}\delta^2(\vec{x}-\vec{y}),
    &
 \{ \phi^i_A(\vec{x},t), \phi^j_H(\vec{y},t) \}
         =
 - \epsilon^{ij}\delta^2(\vec{x}-\vec{y}),\nonumber
    \\
\{\phi^i_V(\vec{x},t) , \phi^j_H(\vec{y},t) \}
         =
     - \epsilon^{ij}\delta^2(\vec{x}-\vec{y}).&
\end{array}
\end{eqnarray}
Therefore the four constraints
$\phi^0_A, \phi^0_H, \Phi_A', \Phi_H'$ are first class and
the eight constraints
$\phi^0_V, \Phi_V, \phi^i_A, \phi^i_V, \phi^i_H$ are second
class.  For the first class constraints
$\phi^0_A, \phi^0_H, \Phi_A', \Phi_H'$, we choose
the gauge fixing conditions:
\begin{eqnarray}
\chi^0_A&\equiv& A^0=0,\quad
\chi_A\equiv
   \frac{1}{2} \partial_iA^i
   -\frac{1}{i\theta_{CS}}\epsilon_{ij}\partial^i(\pi_A^j
   -\pi_V^j) =0,
\cr
\chi^0_H&\equiv& H^0=0,\quad
\chi_H\equiv -\epsilon_{ij}\partial^i\pi_V^j=0.
\end{eqnarray}
The Poisson brackets among the first class constraints and
gauge fixing conditions are evaluated:
\begin{eqnarray}
\{ \chi^0_A(\vec{x},t), \phi^0_A(\vec{y},t) \} &=&
\delta^2(\vec{x}-\vec{y}),
     \cr
\{ \chi^0_H(\vec{x},t), \phi^0_H(\vec{y},t) \} &=&
\delta^2(\vec{x}-\vec{y}),
     \cr
\{ \chi_A(\vec{x},t), \Phi_A'(\vec{y},t) \} &=&
- \partial^2_i\delta^2(\vec{x}-\vec{y}),
     \cr
 \{ \chi_H(\vec{x},t), \Phi_H'(\vec{y},t) \} &=&
-\partial^2_i\delta^2(\vec{x}-\vec{y}),
\end{eqnarray}
and the others are zero. Then we get a set of constraints and
gauge fixing conditions, all of which are second class.

%%%%%%%%%%% Partition Function
Now, we get the master partition function in phase
space as~\cite{Senjanovich}
\begin{eqnarray}
Z_M^{3Leading}
   &=&\int {\cal D}A_{\mu} {\cal D}V_{\mu} {\cal D}H_{\mu}
      {\cal D}\pi_A^{\mu} {\cal D}\pi_V^{\mu} {\cal D}\pi_H^{\mu}
     {}\prod\delta(\phi)\delta(\Phi)
                  \delta(\chi)\nonumber
      \\
     && \times \exp\{ i\int d^3x(\pi_A^{\mu}
        \dot{A}_{\mu} +\pi_V^{\mu}\dot{V}_{\mu}+
        \pi_H^{\mu}\dot{H}_{\mu} -{\cal H}_{3Leading})\},
\end{eqnarray}
where
\begin{eqnarray}
\prod\delta(\phi)\delta(\Phi)
                  \delta(\chi)&\equiv&
      \delta(\phi^0_A )\delta( \phi^0_H )
     \delta(\phi^0_V )\delta(\chi^0_A)\delta(\chi^0_H)
     \delta(\chi_A)\delta(\chi_H)\nonumber\\
&&\times\delta(\Phi_A')\delta(\Phi_H' )
\delta(\Phi'_V )\prod_{i}\delta(\phi^i_A )\delta(\phi^i_V)
\delta(\phi^i_H).
\end{eqnarray}
We can easily perform the integration over $A^0$, $H^0$ and
all the momentum variables
$\pi_A^{\mu},\pi_V^{\mu},\pi_H^{\mu}$.
Thus, we obtain the master partition
function in the configuration space up to the leading
order:
\begin{eqnarray}
Z_M^{3Leading}
    & = &\int {\cal D}A_{\mu} {\cal D}V_{\mu} {\cal D}H_{\mu}
    \delta(\partial_iA^i)\delta(\partial_iH^i)
    \delta(\frac{1}{G}V_0+\epsilon^{ij}\partial_iH_j)
   \nonumber
   \\
   &&\qquad\quad \times\exp
   \{i\int d^3x {\cal L}_{3Leading}[A,H,V]
   \},
   \label{3LinterP}
\end{eqnarray}
where we have exponentiated the constraints
$\Phi_A',\Phi_H'$ and recovered $A_0,H_0$ by identifying the
Lagrange multiplier fields for
$\Phi_A',\Phi_H'$ with $A_0,H_0$ respectively.
We notice that, in the partition function in configuration
space (\ref{3LinterP}), the original Lagrangian (\ref{3L})
is recovered.
\par
{}From the partition function (\ref{3LinterP}), we can show that
Kondo's interpolating Lagrangian
${\cal L}_K$ (\ref{interK}) appears after $V_{\mu}$
integration, while Fradkin-Schaposnik's interpolating
Lagrangian
${\cal L}_{FS}$ (\ref{interFS}) appears after
$A_{\mu}$ integration as follows.
\par
At first, we integrate out the $V_{\mu}$ field.
The integration
can be easily performed and the result is
\begin{eqnarray}
Z_M^{3Leading}&=&
   \int {\cal D}A_{\mu}{\cal D}H_{\mu}
   \delta(\partial_iA^i)\delta(\partial_iH^i)
   \exp\{i\int d^3x{\cal L}_K[A,H] \},
   \\
   {\cal L}_K[A,H]&\equiv&
   -\frac{G}{4}F^{\mu\nu}[H]F_{\mu\nu}[H] +
   \frac{1}{2}\epsilon^{\mu\nu\rho}H_{\mu}F_{\nu\rho}[A]\nonumber
   \\
   &&\qquad
   +\frac{i\theta_{CS}}{2}
\epsilon^{\mu\nu\rho}A_{\mu}\partial_{\nu}A_{\rho}.
\label{3LinterK}
\end{eqnarray}
The Lagrangian ${\cal L}_K$ is the same as (\ref{interK}) up
to the leading order.  Using the interpolating Lagrangian
(\ref{3LinterK}), we can show the equivalence of the
Maxwell-Chern-Simons theory and the self-dual model, as
carried out in ~\cite{Kondo95th2}.
\par
Next, we want to perform the $A_{\mu}$ integration. The
$A_0$  integration gives
\begin{eqnarray}
Z_M^{3Leading}
  &=&\int {\cal D}A_{i} {\cal D}V_{\mu} {\cal D}H_{\mu}
  \delta(\partial_iA^i)\delta(\partial_iH^i)
  \delta(\frac{1}{G}V_0+\epsilon^{ij}\partial_iH_j)\nonumber
  \\
  &&\qquad\quad\times
  \delta(
\epsilon^{ij}\partial_i(i\theta_{CS}A_j+H_j)
        )
  \exp\{ i\int d^3x {\cal L}'\},\label{eq:FS'}\\
{\cal L}'&\equiv&
    \frac{1}{2} \epsilon^{\mu\nu\rho}H_{\mu}F_{\nu\rho}[V]
    + \frac{1}{2G} V^{\mu}V_{\mu}
    + \frac{1}{2} \epsilon^{ij}H_0 F_{ij}[A] \nonumber\\
    &&\qquad - \epsilon^{ij}H_i\partial_0 A_j -
    \frac{i\theta_{CS}}{2} \epsilon^{ij}A_{i}\partial_0
A_j.
\end{eqnarray}
Here $A_i$ field should obey the two constrains:
\begin{equation}
\partial_iA^i=0,\quad
\epsilon^{ij}\partial_iA_j=\frac{i}{\theta_{CS}G}
\epsilon^{ij}\partial_iH_j.
\end{equation}
In order to transform the second inhomogeneous constraint to
the homogeneous one, we shift the field as
\begin{eqnarray}
&&A_i=A_i^{cl} + A'_i,
\cr
&&A^{cl}_i(\vec{x},t)\equiv
\epsilon_{ij}\partial^j\int D(\vec{x}-\vec{y})
\frac{i}{\theta_{CS}G}\epsilon^{ij}\partial_iH_j(\vec{y},t)d\vec{y},
\cr
&&\partial_i^2 D(\vec{x})=\delta^2(\vec{x}).
\end{eqnarray}
As a result, the $A'_i$ field satisfies the homogeneous
constraints:
\begin{equation}
\partial^iA'_i=0,\qquad \epsilon^{ij}\partial_iA'_j=0,
\end{equation}
and the partition function reads
\begin{eqnarray}
Z_M^{3Leading}
   &=&\int {\cal D}A'_{i} {\cal D}V_{\mu} {\cal D}H_{\mu}
  \delta(\partial^iA_i')\delta(\partial_iH^i)
  \delta(\epsilon^{ij}\partial_iA_j' )\nonumber
  \\
  &&\qquad\quad\times
  \delta(\frac{1}{G}V_0+\epsilon^{ij}\partial_iH_j)
  \exp\{ i\int d^3x {\cal L}''\},\label{eq:FS''}
  \\
{\cal L}''&\equiv&
    \frac{1}{2}\epsilon^{\mu\nu\rho}H_{\mu}F_{\nu\rho}[V]
    + \frac{1}{2G} V^{\mu}V_{\mu}
    + \epsilon^{ij}\partial_0(A^{cl}_i+ A'_i) H_j\nonumber\\
    && -
    \frac{1}{i\theta_{CS}} \epsilon^{ij}H_0\partial_iH_j +
    \frac{i\theta_{CS}}{2}
    \epsilon^{ij}\partial_0 (A^{cl}_i + A'_{i})
    (A^{cl}_j + A'_j).
\end{eqnarray}
Moreover we integrate out $A'_{i}$ and obtain
\begin{eqnarray}
Z_M^{3Leading}&=&\int
   {\cal D}V_{\mu}{\cal D}H_{\mu}\delta(\partial_iH^i)
   \delta(\frac{1}{G}V_0+\epsilon^{ij}\partial_iH_j)
   \nonumber\\
   &&\times
   \exp \{ i\int d^3x{\cal L}_{FS}[H,V]\},
   \\
   {\cal L}_{FS}[H,V]&\equiv&
   \frac{i}{2\theta_{CS}}
   \epsilon^{\mu\nu\rho}H_{\mu}\partial_{\nu}H_{\rho}
        + \frac{1}{2} \epsilon^{\mu\nu\rho}F_{\mu\nu}[V]H_{\rho}
        + \frac{1}{2G}V^{\mu}V_{\mu} .
\end{eqnarray}
The Lagrangian ${\cal L}_{FS}$ coincides with the one
(\ref{interFS})
given in ref.~\cite{FS94}.
\par
Finally, we integrate out both $H_{\mu}$ and $V_{\mu}$ fields.
We perform
$H_0$ integration to result in the delta function
$\delta(\epsilon^{ij}F_{ij}[A+V])$.
As in the case of $A_i$ integration of (\ref{eq:FS'}), we
shift the $H_i$ field as
\begin{eqnarray}
   H_i&=&H_i^{cl} + H'_i,
\cr
   H^{cl}_i(\vec{x},t)&\equiv&
        -\epsilon_{ij}\partial^j\int D(\vec{x}-\vec{y})
        \frac{1}{G}V_0(\vec{y},t)d\vec{y}.
\end{eqnarray}
Then the partition function is rewritten as
\begin{eqnarray}
Z_M^{3Leading} &=&
   \int{\cal D}A_{\mu}{\cal D}V_{\mu}{\cal D}H'_{i}
  \delta(\partial_iA^i)\delta(\epsilon^{ij}F_{ij}[A+V])
    \cr   && \times
\delta(\partial_iH'^i)\delta(\epsilon^{ij}\partial_iH'_j)
   \exp\{ i\int d^3x {\cal L}'''\},
     \\
   {\cal L}'''&\equiv&
-\epsilon^{ij}(H^{cl}+H')_i\partial_0(A+V)_j
-\frac{1}{G}V_0(A+V)_0
    \cr &&
+\frac{i\theta_{CS}}{2}
\epsilon^{\mu\nu\rho}A_{\mu}\partial_{\nu}A_{\rho}
+\frac{1}{2G}V^{\mu}V_{\mu}.
\end{eqnarray}
We can transform the integration measure ${\cal D}V_i$ as
\begin{equation}
{\cal D}V_i \delta(\epsilon^{ij}F_{ij}[A+V])
={\cal D}\theta,
\end{equation}
where we have used the solution
$V_i=\sqrt{N}\partial_i\theta -A_i$ for the pure gauge
constraint
$\epsilon^{ij}F_{ij}[A+V]=0$. After the residual
$H'_i$ and
$V_0$ integration, we arrive at
\begin{equation}
   Z_M^{3Leading}
   =\int {\cal D}A_{\mu}{\cal D}\theta\delta(\partial_iA^i)
   \exp\{i\int d^3x{\cal L}_{Th''}[A,\theta]\},
\end{equation}
where ${\cal L}_{Th''}$ is the gauge-invariant Lagrangian
(\ref{Th''}) of the Thirring model in which the
St\"uckelberg field is introduced to recover the gauge
invariance.
\par
If we change the variable $\theta$ to
\begin{equation}
\varphi\equiv
\sqrt{\frac{N}{2G}}e^{i\theta} ,
\end{equation}
then we obtain
\begin{eqnarray}
   Z_M^{3Leading}
   &=&\int {\cal D}A_{\mu}{\cal D}\varphi\delta(\partial_iA^i)
   \exp\{i\int d^3x{\cal L}_{3gNL\sigma}[A,\varphi]\},
\\
{\cal L}_{3gNL\sigma}[A,\varphi]
      &=&(D_{\mu}[A]\varphi)^{\dagger}(D^{\mu}[A]\varphi)
   + \frac{i\theta_{CS}}{2}
      \epsilon^{\mu\nu\rho}A_{\mu}\partial_{\nu}A_{\rho}.
      \label{3DsigmaL}
\end{eqnarray}
This model (\ref{3DsigmaL}) is nothing but the
gauged non-linear $\sigma$ model, if we identify $\theta$ as
a phase variable of the scalar field $\varphi$.
We should remark that the phase variable $\theta$ of the
scalar field $\varphi$ can be divided into two parts, one of
which is the multi-valued function corresponding to the
topologically nontrivial sector and another is a
single-valued function describing the fluctuation around a
given topological sector \cite{KL94}.  In this paper we
take into account the single-valued function only.
Apart from this subtlety, we have shown the equivalence of
the massive Thirring model to the non-linear $\sigma$ model.

%%%%%%%%%%%%%%% 3 dim. Next-To-Leading order
\subsection{up to the next-to-leading order}
Next, we examine the master Lagrangian up to the
next-to-leading order of
$\frac{1}{m}$, because the canonical structure is
different from that of the leading order. Up to
the next-to-leading order, the master Lagrangian is given
by
\begin{eqnarray}
{\cal L}_{3Next}[A,H,V] &=&
   \frac{i\theta_{CS}}{2}
   \epsilon^{\mu\nu\rho}A_{\mu}\partial_{\nu} A_{\rho}
     -\frac{1}{24\pi|m|} F^{\mu\nu}[A]F_{\mu\nu}[A]
     \nonumber    \\
     &&\qquad\quad +  \frac{1}{2G} V_{\mu}V^{\mu}
     +\frac{1}{2}\epsilon^{\mu\nu\rho}
  H_{\mu}F_{\nu\rho}[A+V].
     \label{eq:3DnextLag}
\end{eqnarray}
{}From the Lagrangian, we get the primary constraints:
\begin{eqnarray}
\phi^0_A &\equiv& \pi^0_A \approx  0,\cr
\phi^0_V &\equiv& \pi^0_V \approx  0,\quad
\phi^i_V \equiv  \pi^i_V -\epsilon^{ij}H_j\approx0, \cr
\phi^{\mu}_H &\equiv& \pi^{\mu}_H \approx 0,
\end{eqnarray}
and the canonical Hamiltonian:
\begin{eqnarray}
{\cal H}_{3Next} &=& 3\pi |m|\{\pi^i_A
-\epsilon_{ij}( \frac{i\theta_{CS}}{2}A^j+H_j )\}^2
   + \pi^i_A\partial_iA_0
             \cr
   && +\frac{1}{24\pi|m|}F^{ij}[A]F_{ij}[A]
             -\frac{1}{2G}V^{\mu}V_{\mu}
    -\frac{1}{2}H_0\epsilon^{ij}F_{ij}[A+V]
   \cr
   && -\epsilon^{ij}H_i\partial_jV_0
   -\frac{i\theta_{CS}}{2}\epsilon^{ij}A_0\partial_iA_j
   -3\pi i\theta_{CS}|m|H^iA_i,
\end{eqnarray}
{}From the Hamiltonian, we get the secondary constraints:
\begin{eqnarray}
\Phi_A &\equiv&
      \partial_i(\pi^i_A + \frac
{i\theta_{CS}}{2}\epsilon^{ij}A_j
      )\approx 0,
      \cr
\Phi_V &\equiv&
      \frac{1}{G}V_0 + \epsilon^{ij}\partial_i H_j\approx 0,
      \cr
\Phi_H &\equiv&
     \frac{1}{2} \epsilon^{ij}F_{ij}[A+V]\approx 0.
\end{eqnarray}
It turns out that the canonical structure up to the next-to
leading order is different from that in the leading order,
because all
$\pi^{\mu}_A$'s up to the leading order are constrained.
Following the same steps as in the leading-order case, we
can see that the the four constraints
$\phi^0_A,\Phi_A,\phi^0_H,\Phi_H$ are first class and the
six constraints
$\phi_V^0,\phi^i_V,\Phi_V,\phi_H^i$ are second class. For
these first class constraints, we choose the gauge fixing
conditions:
\begin{eqnarray}
\chi^0_A\equiv A^0 = 0,&&\quad \chi_A\equiv\partial_iA^i=0,
    \cr
\chi^0_H\equiv H^0 = 0,&&\quad
\chi_H\equiv\epsilon_{ij}\partial^i\pi_{V}^{j}=0.
\end{eqnarray}
It can be easily shown that the first class constraints
$\phi^0_A,\Phi_A,\phi^0_H,\Phi_H$ and the gauge fixing
conditions
$\chi^0_A,\chi_A,\chi^0_H,\chi_H$ form a set of second
class constraints. Then, we can obtain the master partition
function in the phase space:
\begin{eqnarray}
Z_M^{3Next} &=& \int {\cal D}A_{\mu}{\cal D}V_{\mu}
       {\cal D}H_{\mu}
         {\cal D}\pi_A^{\mu}{\cal D}\pi_V^{\mu}
         {\cal D}\pi_H^{\mu}
         \prod\delta(\phi)\delta(\Phi)\delta(\chi)\nonumber
         \\
     &&\qquad \times \exp\{ i\int
    d^3x(\pi_A^{\mu}\dot{A}_{\mu}+\pi_V^{\mu}\dot{V}_{\mu}+
        \pi_H^{\mu}\dot{H}_{\mu} -{\cal H}_{3Next})\},
\end{eqnarray}
where
\begin{eqnarray}
\prod\delta(\phi)\delta(\Phi)\delta(\chi)&\equiv&
      \delta(\phi^0_A )\delta( \phi^0_H )
     \delta(\phi^0_V )\delta(\chi^0_A)\delta(\chi^0_H)
     \delta(\chi_A)\delta(\chi_H)\nonumber\\
&&\times\delta(\Phi_A)\delta(\Phi_H )
\delta(\Phi_V )\prod_{i}\delta(\phi^i_V)\delta(\phi^i_H).
\end{eqnarray}
As in the leading-order case, we can perform the
integration over $A_0$, $H_0$ and all the momentum
variables
$\pi^{\mu}_A,\pi^{\mu}_V,\pi^{\mu}_H$, and exponentiate
the constrains $\Phi_A$
and
$\Phi_H$ by identifying the Lagrange multiplier fields
for $\Phi_A,\Phi_H$ with $A_0,H_0$ respectively.
As a result, we obtain the master partition function in
configuration space  up to the next-to-leading
order:
\begin{eqnarray}
Z_M^{3Next} &=& \int {\cal D}A_{\mu}
{\cal D}V_{\mu}{\cal D}H_{\mu}
         \delta(\partial_iA^i)\delta(\partial_iH^i)
         \delta(\frac{1}{G}V_0+\epsilon^{ij}\partial_iH_j)
         \nonumber
         \\
     &&\qquad \times \exp\{ i\int d^3x{\cal L}_{3Next}
     [A,H,V]\}.
\end{eqnarray}

By using the
same method as in the previous section, we can easily get
the partition function with the interpolating Lagrangian of
{}~\cite{Kondo95th2} after $V_{\mu}$ integration:
\begin{eqnarray}
Z_M^{3Next} &=&\int {\cal D}A_{\mu}{\cal D}H_{\mu}
        \delta(\partial_iA^i)\delta(\partial_iH^i)
         \exp\{ i\int d^3x\tilde{\cal L}_{K}[A,H]\}
         \\
\tilde{\cal L}_K[A,H] &\equiv&
       \frac{i\theta_{CS}}{2}
          \epsilon^{\mu\nu\rho}A_{\mu}\partial_{\nu}A_{\rho}
          -\frac{1}{24\pi |m|} F^{\mu\nu}[A]F_{\mu\nu}[A]
      \nonumber\\
          &&\qquad
      +\frac{1}{2}\epsilon^{\mu\nu\rho}H_{\mu}F_{\nu\rho}[A]
      -\frac{G}{4}
F^{\mu\nu}[H]F_{\mu\nu}[H].\label{3LinterK-next}
\end{eqnarray}

On the other hand,  after the $A_{\mu}$ integration,
we obtain the partition function:
\begin{eqnarray}
Z_M^{3Next} &=& \int {\cal D}V_{\mu}{\cal D}H
     \delta(\partial_iH^i)
     \delta(\frac{1}{G}V^0+\epsilon^{ij}\partial_iH_j)
     \nonumber\\
   &&\qquad\times
   \exp\{i\int d^3x \tilde{\cal L}_3[H,V]\},
   \\
\tilde{\cal L}_3[H,V]
   &=&
   + \frac{i}{2\theta_{CS}}
      \epsilon^{\mu\nu\rho}H_{\mu}\partial_{\nu}H_{\rho}
   + \frac{1}{24\pi \theta^2_{CS}|m|}
   F^{\mu\nu}[H]F_{\mu\nu}[H]
   \nonumber\\
   &&\qquad \quad
   + {1 \over 2G} V^\mu V_\mu
   + \frac{1}{2}
      \epsilon^{\mu\nu\rho}H_{\mu}F_{\nu\rho}[V]
      + {\cal O}(\frac{\partial^2}{m^2}).
    \label{3DinterFSnext}
\end{eqnarray}
This is an extension of the interpolating Lagrangian
(\ref{interFS}) into the next-to-leading order.
\par
Finally, we can integrate out both $H_{\mu}$ and $V_{\mu}$
fields as in the leading-order case and get
\begin{equation}
   Z_M^{3Next}
   =\int {\cal D}A_{\mu}{\cal D}\theta\delta(\partial_iA^i)
   \exp\{i\int d^3x{\cal L}_{Th''}[A,\theta]\},
\end{equation}
where ${\cal L}_{Th''}$ is the gauge-invariant Lagrangian
(\ref{Th''}) of the Thirring model up to the
next-to-leading order.
\par
Following the argument given in the previous subsection, the
partition function of the Lagrangian (\ref{eq:3DnextLag}) is identified
with that of the gauged non-linear $\sigma$ model:
\begin{eqnarray}
Z_M^{3Next} &=& \int {\cal D}A_{\mu}{\cal D}
\varphi\delta(\partial_iA^i)
   \exp\{i\int d^3x
   \tilde{\cal L}_{3gNL\sigma}[A,\varphi] \},
   \\
   \tilde{\cal L}_{3gNL\sigma}[A,\varphi]
   &=&(D_{\mu}[A]\varphi)^{\dagger}
   (D^{\mu}[A]\varphi)
   + \frac{i\theta_{CS}}{2}
      \epsilon^{\mu\nu\rho}A_{\mu}\partial_{\nu}A_{\rho}
   \nonumber \\
   &&\qquad \qquad
   - \frac{1}{24\pi |m|}F^{\mu\nu}[A]F_{\mu\nu}[A].
\end{eqnarray}

%%%%%%%%%%%%%%%%%%%% 2 dim. space-time

\section{(1+1) dimensions}
\setcounter{equation}{0}
In 1+1-dimensional case, we can show the equivalence between
the massive Thirring model and a free scalar theory.
\par
The master Lagrangian in 1+1-dimensional space-time is given as
\begin{equation}
{\cal L}_{2}[A,H,V]
         =-\frac{1}{4\pi m^2}F_{\mu\nu}[A]F^{\mu\nu}[A]
         +\frac{1}{2G}V_{\mu}V^{\mu}
         +\frac{1}{2}H\epsilon^{\mu\nu}F_{\mu\nu}[A+V].
\label{eq:2dMaster}
\end{equation}
We should note that the $H$ field is a scalar field in 1+1
dimensional space-time.
{}From the Lagrangian (\ref{eq:2dMaster}), we get the primary
constraints:
\begin{eqnarray}
   \phi^0_A &\equiv& \pi^0_A\approx 0,\cr
   \phi^0_V&\equiv&\pi^0_V\approx 0,\qquad
   \phi^1_V \equiv \pi^1_V-\epsilon^{01}H\approx 0,\cr
   \phi_H &\equiv&\pi_H\approx 0,
\end{eqnarray}
and Hamiltonian:
\begin{equation}
{\cal H}_{2}=\frac{\pi m^2}{2}(\pi_A^1 + \epsilon_{01}H)^2
       + \pi_A^1\partial_1A_0 +\epsilon^{01}H\partial_1V_0
       -\frac{1}{2G}V^{\mu}V_{\mu}.
\end{equation}
{}From the Hamiltonian, we obtain the secondary constraints:
\begin{eqnarray}
  \Phi_A &\equiv& \partial_1\pi^1_A\approx 0,\cr
  \Phi_V &\equiv& \frac{1}{G} V^0
         +\epsilon^{01}\partial_1H\approx 0.
\end{eqnarray}
Calculating the Poisson brackets among these constraints
$\phi^0_A,\phi^0_V,\phi^1_V,\phi_H,\Phi_A$, \break
$\Phi_V$,
we can see the two constraints
$\phi^0_A$,
$\Phi_A$ are first class and
the four constraints $\phi_H,\phi^1_V,\phi^0_V,\Phi_V$ are
second class.
For the first class constraints
$\phi^0_A,\Phi_A$, we choose the gauge fixing conditions:
\begin{equation}
\chi^0_A \equiv A^0 = 0,
      \qquad \chi_A\equiv \partial_1A^1 =0.
\end{equation}
Using these constraints and gauge fixing conditions, we obtain
the master partition function in phase space:
\begin{eqnarray}
Z_M^{2} &=& \int {\cal D}A_{\mu}{\cal D}V_{\mu}{\cal D}H
         {\cal D}\pi_A^{\mu}{\cal D}\pi_V^{\mu}{\cal D}\pi_H
         \prod\delta(\phi) \delta(\Phi)\delta(\chi)\nonumber
        \\
     &&\qquad \times \exp\{
i\int d^2x(\pi_A^{\mu}\dot{A}_{\mu}+\pi_V^{\mu}\dot{V}_{\mu}+
        \pi_H\dot{H} -{\cal H}_{2})\},
\end{eqnarray}
where
\begin{equation}
\prod\delta(\phi) \delta(\Phi)\delta(\chi)
     \equiv \delta(\phi_A^0)\delta(\phi_V^0)\delta(\phi_H)
     \delta(\phi_V^1)\delta(\Phi_A)\delta(\Phi_V)
     \delta(\chi_A^0)\delta(\chi_A).\nonumber
\end{equation}
We can perform $\pi_H$, $\pi_V^0$, $\pi_V^1$ and $A_0$
integration and get the master partition function in
configuration space:
\newpage
\begin{eqnarray}
Z_M^{2}&=&\int {\cal D}A_{\mu}{\cal D}V_{\mu}
    {\cal D}H\delta (\partial_1A^1)
  \delta (\frac{1}{G}V^0+\epsilon^{01}\partial_1H)
  \cr
  &&\qquad\times\exp\{i\int d^2x{\cal L}_{2}[A,H,V]\},
\end{eqnarray}
where we have identified the Lagrange multiplier field for
$\Phi_A$ with $A^0$.
\par
Integrating out $V_{\mu}$ field, we get
\begin{eqnarray}
Z_M^{2}&=&\int{\cal D}A_{\mu}{\cal D}H\delta(\partial_1A^1)
     \exp\{ i\int d^2x {\cal L}_{2K}[A,H]\},\label{kondo2dP}
     \\
{\cal L}_{2K}[A,H] &=&
\frac{G}{2}\partial_{\mu}H\partial^{\mu}H -\frac{1}{4\pi
m^2}F_{\mu\nu}[A]F^{\mu\nu}[A] +\frac{1}{2} H
\epsilon^{\mu\nu}F_{\mu\nu}[A].
\label{eq:kondo2d}
\end{eqnarray}
Indeed, this Lagrangian (\ref{eq:kondo2d}) is identical to
the interpolating Lagrangian introduced in
{}~\cite{Kondo95th2} which shows the equivalence of the
massive Thirring model in 1+1-dimensional space-time to
the scalar theory with the Lagrangian:
\begin{equation}
{\cal L}_{ \rm scalar}[H]=
\frac{1}{2}(G+\frac{6\pi}{5})\partial_{\mu}H\partial^{\mu}H
-3\pi m^2 H^2 + {\cal O}(\frac{1}{m^2}).\label{scalar}
\end{equation}
%%%%%%%%%%% Insert
If we perform the integration over the $H$ field in the
interpolating  Lagrangian
(\ref{eq:kondo2d}), we expect to get the Lagrangian
\begin{equation}
-\frac{1}{4\pi m^2}F_{\mu\nu}[A]F^{\mu\nu}[A]
+\frac{1}{2G}A_\mu^2.\label{2dL[A]}
\end{equation}
Nevertheless, we cannot get this result because the
interpolating partition function (\ref{kondo2dP}) is not
covariant. In fact, we can obtain the result
(\ref{2dL[A]}) corresponding to (\ref{selfdual}) as
mentioned
in introduction if we take the covariant gauge-fixing
condition.
%%%%%%%%%%%% Insert
\par
On the other hand, integrating out the $V_{\mu}$ and $H$
fields, we get the (1+1)-dimensional massive Thirring
model:
\begin{eqnarray}
Z_M^{2}&=&\int{\cal D}A_{\mu} {\cal D}\theta
\delta(\partial_1A^1)
     \exp\{ i\int d^2x {\cal L}_{2Th}[A,\theta ]\}
     \\
{\cal L}_{2Th}[A,\theta ]
      &=&
      \frac{1}{2G}(A_{\mu}-\sqrt{N}\partial_{\mu}\theta)^2
      -\frac{1}{4\pi m^2}F_{\mu\nu}[A]F^{\mu\nu}[A].
\end{eqnarray}
Furthermore, the Lagrangian ${\cal L}_{2Th}$ is
rewritten as
\begin{eqnarray}
&&{\cal L}_{2Th}[A,\theta] ={\cal L}_{2gNL\sigma}[A,\varphi ]
     \cr &&\quad
   \equiv (D_{\mu}[A]\varphi)^{\dagger} (D^{\mu}[A]\varphi)
         -\frac{1}{4\pi m^2}F_{\mu\nu}[A]F^{\mu\nu}[A],
         \\
&&\qquad \varphi \equiv \sqrt{\frac{N}{2G}}e^{i\theta},
   \nonumber
\end{eqnarray}
as well as in the (2+1)-dimensional case. This Lagrangian
is that of the gauged non-linear $\sigma$ model.

%%%%%%%%%%%%%%%%%%%% Higher dim. -> Reducible Constraints

\section{(d+1) dimensions }
\setcounter{equation}{0}
We want to perform the same procedure for $D=d+1\ge 4$ case,
but  in this case there appears the reducible constraint.
The master Lagrangian in $D$ dimensional space-time is
given by
\begin{equation}
{\cal L}_{D}[A,H,V] = {\cal L}_G[A]
          + \frac{1}{2G}V^{\mu}V_{\mu}
          + \frac{1}{2}
          \epsilon^{\mu_1 ... \mu_D} H_{\mu_3 ... \mu_D}
      F_{\mu_1\mu_2}[A+V],\label{D-dimLag}
\end{equation}
where $H_{\mu_3 ... \mu_D}$ is a totally
anti-symmetric tensor field of rank $D-2$.
Note that
\begin{eqnarray}
 {\cal L}_G[A] &=&
-\frac{\kappa_D}{4}F^{\mu\nu}[A]F_{\mu\nu}[A],
\end{eqnarray}
where $\kappa_D$ is a divergent
constant which depends on regularization-scheme and
dimensionality $D$, see e.g.~\cite{Kondo95th2}.
{}From the Lagrangian (\ref{D-dimLag}), we get
the constraints
\begin{equation}
\Phi^{i_1 ... i_{D-3}}_H \equiv
\epsilon^{i_1 ... i_{D-1}}F_{i_{D-2}i_{D-1}}[A+V]\approx 0,
\end{equation}
as secondary constraints. However, the constraints
are not independent, because they satisfy the relations
\begin{eqnarray}
F_1^{i_1 ... i_{D-4}}&\equiv&
\partial_{i_{D-3}}\Phi^{{i_1 ...i_{D-3}}}_H =
0,\label{D-reducible}
\end{eqnarray}
where we should note that the relations are identically zero.
By the same logic as above, the relations
(\ref{D-reducible}) are not independent each other.
Therefore the theory with the  master Lagrangian
(\ref{D-dimLag}) is a reducible theory of $(D-3)$-th stage,
irrespective of the explicit form of ${\cal L}_G[A]$. Even
in the reducible theory we can repeat the same treatment as
in the previous sections. However, there is no guarantee
that the covariant theory is obtained as the final result.
If we want to get the covariant result without failure, we
must resort to other method, for example, BFV
method~\cite{BFV}.

%%%%%%%%%%%%%%%%%%%%%%%%%%%%%%%%%%%
\subsection{(3+1) dimensions}

{}From the above reason, we treat the (3+1) dimensional system
based on the BFV method.

The master Lagrangian in
$3+1$ dimensional space-time is given by
\begin{eqnarray}
{\cal L}_{4}[A,H,V] &=& {\cal L}_G[A]
          + \frac{1}{2G}V^{\mu}V_{\mu}
          + \frac{1}{2}
          \epsilon^{\mu\nu\rho\sigma} H_{\mu\nu}
      F_{\rho\sigma}[A+V],\label{eq:4dimLag}
       \\
 {\cal L}_G[A] &=&
     -\frac{\kappa}{4}F^{\mu\nu}[A]F_{\mu\nu}[A],
\end{eqnarray}
where $H_{\mu\nu}$ is an anti-symmetric tensor field and
$\kappa$
is a regularization-scheme-dependent divergent constant.
{}From the Lagrangian ${\cal L}_4$, we get the primary
constraints
\begin{eqnarray}
\phi_A^0   &\equiv& \pi_A^0\approx 0,\cr
\phi_V^0 &\equiv& \pi_V^0\approx 0,\qquad
\phi_V^i \equiv \pi_V^i-\epsilon^{ijk}H_{jk}\approx 0,\cr
\phi^{0i}_H &\equiv& \pi_H^{0i}\approx 0,\qquad
\phi^{ij}_H\equiv \pi_H^{ij}\approx 0.\label{PC}
\end{eqnarray}
and
the Hamiltonian
\begin{eqnarray}
{\cal H}&=&\int d^3x\{ -\frac{1}{2\kappa}
       (\pi^i_A - \epsilon^{ijk}H_{jk})(\pi_{Ai}
       -\epsilon_{ilm}H^{lm})
        -A_0\partial_i\pi_A^i
        \cr &&
      + \frac{\kappa}{4}F_{ij}[A]F^{ij}[A]
      -\epsilon^{ijk}H_{0i}F_{jk}[A+V]
      -\epsilon^{ijk}V_0\partial_iH_{jk}
      -\frac{1}{2G}V^{\mu}V_{\mu}\}.\label{4DH}
\end{eqnarray}
For the constraints $\phi^0_A$ and $\phi^{0i}_H$, we choose
the gauge-fixing conditions
\begin{equation}
A_0=H_{0i}=0.
\end{equation}
So, we eliminate the variables
$(\pi^0_A,A_0)$ and $(\pi^{0i}_H,H_{0i})$ from now on.
The non-vanishing Poisson bracket among the residual
constraints is
\begin{equation}
\{ \phi_H^{ij}(\vec{x},t),\phi_V^k(\vec{y},t) \}
=\epsilon^{ijk}\delta^3(\vec{x}-\vec{y}).
\end{equation}
Therefore, the constraints
$\phi_V^i$ and $\phi^{ij}_H$ are second class. Moreover we
shall eliminate these two second class constraints by solving
them, and the Poisson bracket changes to the modified one:
\begin{equation}
\{ H_{ij},V_k \} = \frac{1}{2} \epsilon_{ijk},
\end{equation}
i.e., we can consider that
$\epsilon^{ijk}H_{jk}$ and $V_i$ form a canonical pair.
{}From the Hamiltonian (\ref{4DH}) we get the secondary
constraints
\begin{eqnarray}
\Phi_A  &\equiv& \partial_i\pi_A^i\approx 0,\cr
\Phi_V  &\equiv& \frac{1}{G}V^0
         + \epsilon^{ijk}\partial_i H_{jk}  \approx 0,\cr
\Phi^i_H  &\equiv& \epsilon^{ijk}F_{jk}[A+V] \approx 0.
           \label{SC}
\end{eqnarray}
The non-vanishing Poisson bracket among the residual
constraints
$\phi_V^0$, $\Phi_A$, $\Phi_V$,
$\Phi^i_H$ is
\begin{eqnarray}
\{ \Phi_V(\vec{x},t),\phi_V^0(\vec{y},t) \}
=\frac{1}{G}\delta^3(\vec{x}-\vec{y}).
\end{eqnarray}
Therefore, the constraints
$\phi_V^0$, $\Phi_V$ are second class constraints and
$\Phi_A$, $\Phi^i_H$ are
first class constraints.
In order to treat second class constraints
$\phi_V^0$, $\Phi_V$,
we shall use the Dirac bracket
\begin{equation}
\{ A,B \}_D \equiv \{ A,B\}
-\{ A,\phi^0_V\} \{ \phi^0_V,\Phi_V\}^{-1}
\{ \Phi_V ,B\}
-\{ A,\Phi_V \} \{ \Phi_V ,\phi^0_V\}^{-1}
\{ \phi^0_V,B\},
\end{equation}
where $A,B$ are arbitrary variables. Then, it turns out that
$(\pi^0_V, V_0)$ is not a canonical pair and not dynamical,
because the Dirac bracket between them is zero:
\begin{equation}
\{ \pi^0_V,  V_0\}_D = 0.
\end{equation}
{}From this fact, we concentrate on the first class
constraints
$\Phi_A$ and $\Phi_H^i$ in the phase space
$(\pi_A^i,A_i)$, $(\epsilon^{ijk}H_{jk},V_i)$.
However, the first class constraints are not
independent, because the constraints
$\Phi^i_H$'s satisfy the relation
\begin{equation}
\partial_{i}\Phi^{i}_H =0,\label{reducible}
\end{equation}
where we note that the relation, i.e., Bianchi identity, is
identically zero. In order to get the Lorentz-covariant
and locally well-defined quantized action, we should quantize
the system without solving the relation  (\ref{reducible}).
This can be done when we quantize the system according to the
BFV method. So we deal with the system in BFV formalism.

In the BFV formalism, we have to prepare the phase space
which consists of the original phase space
\begin{eqnarray}
(\pi_A^i,A_i),\quad (\epsilon^{ijk}H_{jk},V_i),\label{OPS}
\end{eqnarray}
and the extended phase space
\begin{eqnarray}
\begin{array}{@{\,}lllll@{\,}}
 & (B,N), & (\overline{P},C), & (\overline{C},P),
      &{\rm for \ \Phi_A\approx 0},
\\
{\rm Grassmann \  parity}
    &\ \  0\quad 0 &\ \  1\ \   1 &\ \  1\ \   1
\\
{\rm ghost \  number}
    &\ \  0\quad 0 &    -1\ \   1 &    -1\ \   1
\\
 & (B^i,N_i), & (\overline{P}^i,C_i), &(\overline{C}^i,P_i),
      &{\rm for \ \Phi_H^i\approx 0}.
\\
{\rm Grassmann \  parity}
    &\ \   0\quad 0 &\ \ 1\quad  1 &\ \   1\quad 1
\\
{\rm ghost \  number}
    &\ \   0\quad 0 &    -1\ \ \ 1 &     -1\ \ \ 1
\label{EPS}
\end{array}
\end{eqnarray}
%%% reducible constraints %%%%%
However, the constraints $\Phi^i_H$'s satisfy the relation
$\partial_i\Phi^i_H= 0$.
In this case, we must impose the fermionic constraint
\begin{equation}
\partial_i\overline{P}^i\approx 0,\label{RC}
\end{equation}
and accordingly introduce the canonical pairs
\begin{eqnarray}
\begin{array}{@{\,}llll@{\,}}
 & (B_{(1)},N_{(1)}),
&  (\overline{P}_{(1)},C_{(1)}),
&  (\overline{C}_{(1)},P_{(1)}).
\\
{\rm Grassmann \  parity}
    &\ \  1\qquad    1 &\ \  0\qquad 0 &\ \  0\qquad 0
\\
{\rm ghost \  number}
    &    -1\quad \ \ 1 &    -2\qquad 2 &    -2\qquad 2
\label{EPS-RC}
\end{array}
\end{eqnarray}
This is how to treat the
reducible constraints in BFV formalism.  To make the Lorentz
covariance manifest and perform gauge-fixing of the field
$\ol{C}_{\mu}$,  we also introduce the canonical variables
\begin{eqnarray}
\begin{array}{@{\,}lll@{\,}}
 & (B^1_1,N^1_1), & (\overline{C}^1_1,P^1_1),\label{EG}
\\
{\rm Grassmann \  parity}
    &\ \  0\quad 0 &\ \  1\quad 1
\\
{\rm ghost \  number}
    &\ \  0\quad 0 &    -1\ \ \ 1
\end{array}
\end{eqnarray}
as extra ghost fields.
%%%%%%% BRST charge %%%%%%%%
Now, we can construct the BRST charge
\begin{equation}
Q_{BRST}=\int d^3x \{
  C\Phi_A + C_i\Phi_H^i + iC_{(1)}\partial_i\overline{P}^i
  +PB +P_iB^i + P_{(1)}B_{(1)} + P^1_1B^1_1\}.
\end{equation}
{}From the BRST charge, we can calculate the BRST
transformation:
\begin{eqnarray}
\begin{array}{@{\,}ll@{\,}}
\delta A_i=\partial_iC, &
\delta\pi^i_A=2\epsilon^{ijk}\partial_jC_k,
\cr
\delta H_{ij}=\partial_iC_j-\partial_jC_i, &
\cr
\delta V_i = 0, & \delta V_0 =
\delta(-G\epsilon^{ijk}\partial_iH_{jk})=0,
\end{array}
\end{eqnarray}
for the variables
$A_i, \pi_A^i, H_{ij}, V_{\mu}$
and
\begin{eqnarray}
\begin{array}{@{\,}llll@{\,}}
\delta B=0,  &
\delta B^i=0,  &
\delta B_{(1)}=0, &
\delta B^1_1=0,
\cr
\delta N = -P, &
\delta N_i = -P_i, &
\delta N_{(1)} = P_{(1)}, &
\delta N^1_1 = -P^1_1,
\cr
\delta\overline{P}=\Phi_A,&
\delta\overline{P}^i=\Phi_H^i,&
\delta\overline{P}_{(1)}=i\partial_i\overline{P}^i,&
\cr
\delta C =0,   &
\delta C_i = -i\partial_i C_{(1)},  &
\delta C_{(1)} = 0, &
\cr
\delta \overline{C}= B,&
\delta \overline{C}^i= B^i,&
\delta \overline{C}_{(1)}= B_{(1)},&
\delta \overline{C}^1_1= B^1_1,
\cr
\delta P =0, &
\delta P_i =0, &
\delta P_{(1)} =0, &
\delta P^1_1 =0,
\end{array}
\end{eqnarray}
for the residual variables (\ref{EPS}),(\ref{EPS-RC}) and
(\ref{EG}).
Furthermore, we take the gauge-fixing fermion as
\begin{equation}
\Psi = \int d^3x \{ \overline{C}\chi + \overline{P}N
                  + \overline{C}^i\chi_i + \overline{P}^iN_i
                  + \overline{C}_{(1)}\chi_{(1)}
                  + \overline{P}_{(1)}N_{(1)}
                  + \overline{C}^1_1\chi^1_1\},
\end{equation}
where $\chi,\chi_i,\chi_{(1)},\chi^1_1$ are gauge-fixing
functions chosen later. Then, we can give the BRST invariant
quantum action
\begin{eqnarray}
S_q &=&\int d^4x [
     \pi^i_A\dot{A}_i + \epsilon^{ijk}H_{jk}\dot{V}_i
     + B\dot{N} + B^i\dot{N}_i + B_{(1)}\dot{N}_{(1)}
     + B^1_1\dot{N}^1_1
     + \overline{P}\dot{C} + \overline{P}^i\dot{C}_i
     \cr &&\quad
     + \overline{P}_{(1)}\dot{C}_{(1)}
     + \overline{C}\dot{P} + \overline{C}^i\dot{P}_i
     + \overline{C}_{(1)}\dot{P}_{(1)}
     + \overline{C}^1_1\dot{P}^1_1
     - {\cal H}_c -\{\Psi,Q\}_D],
\end{eqnarray}
where ${\cal H}_c$ is the canonical Hamiltonian
\begin{eqnarray}
{\cal H}_c &=& -\frac{1}{2\kappa}
       (\pi^i_A - \epsilon^{ijk}H_{jk})(\pi_{Ai}
       -\epsilon_{ilm}H^{lm})
        + \frac{\kappa}{4}F_{ij}[A]F^{ij}[A]
        \cr && \qquad
      - \epsilon^{ijk}V_0\partial_iH_{jk}
      -\frac{1}{2G}V^{\mu}V_{\mu}.
\end{eqnarray}
Then we can get the master partition function in extended phase space
\begin{eqnarray}
&& \qquad\qquad Z= \int {\cal D}\mu \exp \{ iS_q\},
       \cr
{\cal D}\mu &\equiv &
     {\cal D}A_i    {\cal D}H_{ij}    {\cal D}V_{\mu}
     {\cal D}\pi_A^i
     {\cal D}B      {\cal D}N         {\cal D}\ol{P}
     {\cal D}C      {\cal D}{\ol C}   {\cal D}P
     \cr &&
     \times
     {\cal D}B^i    {\cal D}N_i       {\cal D}\ol{P}^i
     {\cal D}C_i    {\cal D}\ol{C}^i  {\cal D}P_i
     {\cal D}B_{(1)}{\cal D}N_{(1)}   {\cal D}\ol{P}_{(1)}
     {\cal D}C_{(1)} {\cal D}\ol{C}_{(1)} {\cal D}P_{(1)}
     \cr &&
     \times
     {\cal D}B^1_1        {\cal D}N^1_1
     {\cal D}\ol{C}^1_1   {\cal D}P^1_1.
\end{eqnarray}
After we integrate the fields
$\pi_A^i$, $\ol{P}$, $P,\ol{P}^i$,
$P_i$, $\ol{P}_{(1)},P_{(1)}$
and choose the gauge fixing functions
$\chi,\chi_i,\chi_{(1)},\chi_1^1$ as
\begin{eqnarray}
\chi  &=& \partial^iA_i -\frac{\alpha}{2}B,
      \cr
\chi_i &=&
   \partial^jH_{ji}-\partial_iN^1_1-\frac{\beta}{2}B_i,
      \cr
\chi_{(1)} &=& -i\partial^iC_i -i\gamma P^1_1,
      \cr
\chi^1_1   &=& \partial^iN_i -\frac{\beta}{2}B_1^1,
\end{eqnarray}
the partition function in configuration space is expressed as
\begin{eqnarray}
Z &=& \int  {\cal D}A_{\mu}    {\cal D}H_{\mu\nu} {\cal D}V_{\mu}
     {\cal D}B {\cal D}C {\cal D}{\ol C}
     {\cal D}B^{\mu} {\cal D}C_{\mu} {\cal D}{\ol C}^{\mu}
     {\cal D}B_{(1)} {\cal D}N^1_{1} {\cal D}P_1^{1}
     {\cal D}\ol{C}_{(1)}{\cal D}C_{(1)}
     \cr &&\qquad\qquad
     \times \exp\{ i\int d^4x {\cal L}'_q\},
     \\
{\cal L}'_q &\equiv & {\cal L}_{4}[A,H,V]
     -B(\partial^{\mu}A_{\mu}-\frac{\alpha}{2}B)
     +\overline{C}\partial^{\mu}\partial_{\mu}C
       \cr &&\qquad
     -B^{\mu}(\partial^{\nu}H_{\nu\mu}-\partial_{\mu}N^1_1
                       - \frac{\beta}{2}B_{\mu})
     +\overline{C}^{\mu}\partial^{\nu}
         (\partial_{\nu}C_{\mu}-\partial_{\mu}C_{\nu}
          +\partial_\mu P^1_1)
       \cr &&\qquad
       +iB_{(1)}(\partial^{\mu}C_{\mu}
       - \gamma P^1_1)
       + \ol{C}_{(1)}\partial^\mu\partial_{\mu} C_{(1)},
       \label{4DSq}
\end{eqnarray}
where $\alpha,\beta,\gamma$ are gauge-fixing parameters and
we have identified the fields
$N$, $N_i$, $N_{(1)}$, $B^1_1$, $\overline{C}^1_1$
with
$-A_0,-H_{0i},-iC_0,B^0,\overline{C}^0$
respectively to recover the Lorentz covariance.
%%%%%%%
\footnote{
Total Lagrangian ${\cal L}'_q$ can be written in the BRST
invariant form
\begin{eqnarray}
{\cal L}'_q &=& {\cal L}_4 + \delta F,\cr
F &\equiv &
      -\ol{C}(\partial^{\mu}A_{\mu}-\frac{\alpha}{2}B)
      -\ol{C}^{\mu}(\partial^{\nu}H_{\nu\mu}
          -\partial_{\mu}N^1_1 - \frac{\beta}{2}B_{\mu})
      +i\ol{C}_{(1)}(\partial^{\mu}C_{\mu}
          - \gamma P^1_1),
\end{eqnarray}
according to~\cite{HKO}. Here, the covariant BRST
transformation $\delta$ is defined by
\begin{eqnarray}
\begin{array}{@{\,}llll@{\,}}
\delta A_{\mu}=\partial_{\mu} C, &
\delta H_{\mu\nu}
        =\partial_{\mu}C_{\nu}-\partial_{\nu}C_{\mu},&
\delta V_{\mu} = 0,&
\cr
\delta B = 0,  &
\delta B^{\mu} = 0, &
\delta B_{(1)}=0,  &
\delta B^1_1=0,
\cr
\delta C = 0, &
\delta C_{\mu} = -i\partial_{\mu} C_{(1)},&
\delta C_{(1)} = 0, &
\cr
\delta \overline{C}= B, &
\delta \overline{C}^{\mu}= B^{\mu},&
\delta\overline{C}_{(1)} = B_{(1)}, &
\delta \overline{C}^1_1  = B^1_1,
\cr
\delta P^1_1 =0, &
\delta N^1_1 = -P^1_1.
\end{array}
\end{eqnarray}
}
%%%%%%%%%%%%%%%%%%
After all, the partition function results in
\begin{eqnarray}
&&
 Z =\int {\cal D}A_{\mu}{\cal D}H_{\mu\nu}{\cal D}V_{\mu}
        {\cal D}B{\cal D}B^{\mu}{\cal D}N^1_1\
        \exp\{ i{\tilde S}_q +i\tilde{S}_{GF}\},
        \label{4Dpartition}
       \\
&&
 \tilde{S}_q[A,H,V] \equiv  \int d^4x  {\cal L}_4[A,H,V],
     \label{4DQA}
       \\
&&
 \tilde{S}_{GF} \equiv  \int d^4x\{
       -B(\partial^{\mu}A_{\mu}-\frac{\alpha}{2}B)
       -B^{\nu}(\partial^{\mu}H_{\mu\nu}-\partial_{\nu}N^1_1
       -\frac{\beta}{2}B_{\nu})\},
\end{eqnarray}
where we have integrated out the fields
$B_{(1)},C,\overline{C},C_{\mu},\overline{C}^{\mu},P^1_1$.
Besides, we should note that the field $N^1_1$ is still
necessary to fix the gauge degrees of freedom for
$B_{\mu}$. Using the master partition function
(\ref{4Dpartition}) in configuration space, we show the
equivalence among the various theories as follows.

%%%%%%%%%%%%%%%%%%%%%%%%%%%%%%%%%%%%%%%
\subsection{${\cal L}[A,H],\ {\cal L}[A]$}
At first, we integrate out the $V_{\mu}$ field in
(\ref{4Dpartition}). Then the resulting partition function is
\begin{eqnarray}
&&Z=\int {\cal D}A_{\mu}{\cal D}H_{\mu\nu}{\cal D}B
          {\cal D}B_{\mu}{\cal D}N^1_1
     \exp\{ i\int d^4x {\cal L}_4[A,H]+ i\tilde{S}_{GF}\},
 \label{4DP[A,H]}\\
&&{\cal L}_4[A,H] \equiv
    -\frac{\kappa}{4}F^{\mu\nu}[A]F_{\mu\nu}[A]
    +\frac{1}{2}\epsilon^{\mu\nu\rho\sigma}
               H_{\mu\nu}F_{\rho\sigma}[A]
 +\frac{G}{12}\tilde{H}^{\mu\nu\rho}\tilde{H}_{\mu\nu\rho},
      \label{4DL[A,H]} \\
&&{\tilde H}^{\mu\nu\rho}\equiv  \partial^{\mu}H^{\nu\rho}
        +\partial^{\nu}H^{\rho\mu}+\partial^{\rho}H^{\mu\nu}
        -\partial^{\nu}H^{\mu\rho}-\partial^{\mu}H^{\rho\nu}
        -\partial^{\rho}H^{\nu\mu}.\label{Hdef}
\end{eqnarray}
This is an interpolating gauge theory of 3+1 dimensions
corresponding to (\ref{3LinterK}) or (\ref{3LinterK-next}) of
2+1 dimensions.

Furthermore, if $B_{\mu}$, $N^1_1$ and $H_{\mu\nu}$ fields
are integrated out, we get the partition function
\begin{eqnarray}
 Z &=& \int {\cal D}A_{\mu}{\cal D}B
          \exp\{ i\int d^4x {\cal L}_4[A,B]\},
 \\
{\cal L}_4[A,B] &\equiv &
    -\frac{\kappa}{4}F^{\mu\nu}[A]F_{\mu\nu}[A]
    +\frac{1}{4G}F^{\mu\nu}[A]
    \frac{1}{\partial^{\rho}\partial_{\rho}}F_{\mu\nu}[A]
    \cr && \quad
    -B(\partial^{\mu}A_{\mu}-\frac{\alpha}{2}B).
    \label{nextresult}
\end{eqnarray}
This is a gauge theory for the $A_{\mu}$ field with
non-local term. Nevertheless, if we choose the gauge-fixing
parameter
$\alpha=0$,
then the non-local term disappears after the integration of
the $B$ field
and the partition function changes to
\begin{eqnarray}
Z &=&\int {\cal D}A_{\mu} \delta(\partial^{\mu}A_{\mu})
    \exp\{ i\int d^4x {\cal L}_4[A]\},
 \\
{\cal L}_4[A] &\equiv &
    -\frac{\kappa}{4}F^{\mu\nu}[A]F_{\mu\nu}[A]
    -\frac{1}{2G}A_{\mu}A^{\mu}.\label{L[A]}
\end{eqnarray}
This denotes a massive vector theory corresponding to the
self-dual model (\ref{SD}) in (2+1) dimensions.

%%%%%%%%%%%%%%%%%%%%%%%%%%%%%%%
\subsection{${\cal L}[A,\varphi]$}

On the other hand, if we take $\beta = 0$ in
(\ref{4Dpartition}), the $B_{\mu}$, $N^1_1$ and $H_{\mu\nu}$
integrations give
\begin{eqnarray}
&&Z =\int {\cal D}A_{\mu}{\cal D}V_{\mu}{\cal D}B
      \delta(F_{\mu\nu}[A+V])\exp\{ i\int d^4x
      {\cal L}_4[A,V,B]\},
      \\
&&{\cal L}_4[A,V,B]
   \equiv  -\frac{\kappa}{4}F_{\mu\nu}[A]F^{\mu\nu}[A]
      + \frac{1}{2G}V^{\mu}V_{\mu}
      - B(\partial^{\mu}A_{\mu}-\frac{\alpha}{2}B).
\end{eqnarray}
Moreover, we perform the $V_{\mu}$ integration, and obtain
the partition function
\begin{eqnarray}
&& Z =\int {\cal D}A_{\mu}{\cal D}\varphi{\cal D}B
          \exp\{ i\int d^4x L[A,\varphi,B]\},
 \\
&&{\cal L}[A,\varphi,B] \equiv
    -\frac{\kappa}{4}F^{\mu\nu}[A]F_{\mu\nu}[A]
    +\frac{1}{2G}\{
    (\partial^{\mu}-iA^{\mu})\varphi\}^{\dagger}
    \{ (\partial^{\mu}-iA^{\mu})\varphi\}
    \cr &&\qquad\qquad\qquad\qquad
    -B(\partial^{\mu}A_{\mu} - \frac{\alpha}{2}B),
    \\
  &&\qquad\qquad\qquad\qquad
       \varphi \varphi^{*} = 1.\nonumber
\end{eqnarray}
This is a gauged non-linear $\sigma$ model.

%%%%%%%%%%%%%%%%%%%%%%%%%%%%%%%%%%%%%%%%%
\newpage
\subsection{${\cal L}[H,V]$}

Next, the integration of the $A_{\mu}$ field in
(\ref{4Dpartition}) gives the partition function
\begin{eqnarray}
&& Z = \int {\cal D}H_{\mu\nu}{\cal D}V_{\mu}
       {\cal D}B_{\mu} {\cal D}N^1_1
        \exp\{ i\int d^4x {\cal L}[H,V,B_{\mu},N^1_1]\},
  \\
 && {\cal L}[H,V,B_{\mu},N^1_1] \equiv
     \frac{1}{12\kappa}\tilde{H}^{\mu\nu\rho}
     \frac{1}{\partial^{\sigma}\partial_{\sigma}}
     \tilde{H}_{\mu\nu\rho}
     +\frac{1}{2G}V^{\mu}V_{\mu}
     +\frac{1}{2}\epsilon^{\mu\nu\rho\sigma}
     H_{\mu\nu}F_{\rho\sigma}[V]
     \cr &&\qquad\qquad\qquad\qquad
     -B^{\nu}(\partial^{\mu}H_{\mu\nu}-\partial_{\nu}N^1_1
     -\frac{\beta}{2}B_{\nu})
     .\label{finalresult}
\end{eqnarray}
This result is $\alpha$-independent.
If we take the gauge-fixing parameter $\beta =0$, we get
\begin{eqnarray}
Z &=& \int {\cal D}H_{\mu\nu}{\cal D}V_{\mu}
        \delta(\partial^{\mu}H_{\mu\nu})
        \exp \{ i\int d^4x {\cal L}[H,V]\},
        \\
{\cal L}[H,V] &\equiv &
    -\frac{1}{\kappa}H^{\mu\nu}H_{\mu\nu}
       +\frac{1}{2G}V^{\mu}V_{\mu} +
       \frac{1}{2}\epsilon^{\mu\nu\rho\sigma}
       H_{\mu\nu}F_{\rho\sigma}[V],\label{4DL[H,V]}
\end{eqnarray}
after $B_{\mu}$ and $N^1_1$ integrations. It should be
remarked that the non-local term for
$H_{\mu\nu}$
changes into a mass term only if we choose $\beta =0$.

%%%%%%%%%%%%%%%%%%%%%%%%%%
\subsection{${\cal L}[H]$}

Finally, integration of $V_{\mu}$ in (\ref{finalresult})
leads to
\begin{eqnarray}
&& Z= \int {\cal D}H_{\mu\nu}{\cal D}B_{\mu}{\cal D}N^1_1
        \delta(\partial^{\mu}H_{\mu\nu})
        \exp \{ i\int d^4x {\cal L}[H,B_{\mu},N^1_1]\},
        \\
&& {\cal L}[H,B_{\mu},N^1_1] \equiv
    \frac{G}{12}\tilde{H}^{\mu\nu\rho}
     \tilde{H}_{\mu\nu\rho}
     +\frac{1}{12\kappa}\tilde{H}^{\mu\nu\rho}
     \frac{1}{\partial^{\sigma}\partial_{\sigma}}
     \tilde{H}_{\mu\nu\rho}
     \cr &&\qquad\qquad\qquad\qquad
     -B^{\nu}(\partial^{\mu}H_{\mu\nu}-\partial_{\nu}N^1_1
     -\frac{\beta}{2}B_{\nu}).\label{L[H]1}
\end{eqnarray}
If we take $\beta =0$, we get
\begin{eqnarray}
Z&=& \int {\cal D}H_{\mu\nu}
        \delta(\partial^{\mu}H_{\mu\nu})
        \exp \{ i\int d^4x {\cal L}[H]\},
        \\
{\cal L}[H] &\equiv & \frac{G}{12}\tilde{H}^{\mu\nu\rho}
     \tilde{H}_{\mu\nu\rho}
     -\frac{1}{\kappa}H^{\mu\nu}H_{\mu\nu},\label{L[H]2}
\end{eqnarray}
by the integration of $B_{\mu}$ field. As mensioned above,
the non-local term changes to a mass term in the case
$\beta =0$.
After all, we have bosonized the massive Thirring model to
get a tensor gauge theory. This result is consistent with
\cite{Cortes95}. The result (\ref{L[H]2}) can be also
obtained from the partition function (\ref{4DP[A,H]}) by
performing the integrations over all the fields except
$H_{\mu\nu}$.
\par
In the general $D$-dimensional case ($D \ge 4$), it is
expected that the bosonized theory is given by
\begin{eqnarray}
Z&=& \int {\cal D}H_{\mu_1...\mu_{D-2}}
        \delta(\partial^{\mu_1}H_{\mu1...\mu_{D-2}})
        \exp \{ i\int d^4x {\cal L}_D[H]\},
        \\
{\cal L}_D[H] &\equiv &
     \frac{G}{12}\tilde{H}^{\mu_1...\mu_{D-1}}
     \tilde{H}_{\mu_1...\mu_{D-1}}
    -\frac{1}{\kappa_D}
H^{\mu_1...\mu_{D-2}}H_{\mu_1...\mu_{D-2}},\label{D-dimL[H]}
\end{eqnarray}
where
$
\tilde{H}^{\mu_1...\mu_{D-1}}
$
is totally anti-symmetrized one of
$
\partial^{\mu_1}H^{\mu_2...\mu_{D-1}}
$
, i.e.,
\begin{eqnarray}
\tilde{H}^{\mu_1...\mu_{D-1}}
&= & (D-2)!\  (
     \partial^{\mu_1}H^{\mu_2\mu_3...\mu_{D-1}}
       -\partial^{\mu_2}H^{\mu_1\mu_3...\mu_{D-1}}
       +\cdots \cr &&\qquad\qquad\qquad\qquad\qquad
   +(-1)^{(D-2)}\partial^{\mu_{D-1}}H^{\mu_1...\mu_{D-2}}).
\end{eqnarray}

\section{Conclusion and Discussion}
\setcounter{equation}{0}
In this paper we have proposed a master Lagrangian
(\ref{master1}) or (\ref{master2}) for  performing the
bosonization of the Thirring model in arbitrary dimension.
Especially, in (2+1) dimensions, this master Lagrangian is
able to interpolate the previous two interpolating
Lagrangians
\cite{FS94,Kondo95th2}. Starting from the phase-space path
integral formulation of the gauge theory defined by the
master Lagrangian, we have shown the equivalence of the
(2+1)-dimensional massive Thirring model with the
Maxwell-Chern-Simons theory, up to the next-to-leading order
of  $1/m$.
Incidentally it is not difficult to show the
equivalence by applying the generalized canonical formalism
of Batalin, Fradkin, Vilkovisky and Tyutin \cite{BF} to
our master Lagrangian, as carried out in the recent work
\cite{BR95} for the self-dual model.
%%%%%%%%% for 3+1
Actually, in (3+1)-dimensional case, we have shown based on
the BFV method that the bosonized theory of the massive
Thirring model is equivalent to the massive antisymmetric
tensor theory.

Athough the Thirring model in $D>2$ dimensions is
perturbatively nonrenormalizable, the bosonization technique
may throw light on the nonperturbative renormalizability of
the Thirring modelin (3+1)-dimensions as the normalizability
of the four-fermion interaction in $1/N_f$
expansion.~\cite{KSY91}~\cite{HKKN94}

\par
The most interesting question will be how to generalize
our strategy of bosonization into the non-Abelian case.
First we remark that it is easy to show the classical
equivalence of the non-Abelian-gauged non-linear
$\sigma$-model with the gauge-invariant formulation of the
non-Abelian Thirring model with the St\"uckelberg field
\cite{FMMY81}.  Indeed the non-Abelian version of our master
Lagrangian is at least classically equivalent to the
non-Abelian-gauged non-linear $\sigma$-model
\cite{Townsend79,FT81,TM90}.  In this sense our master
Lagrangian is easily extended to the non-Abelian case.
However, we must specify the gauge-fixing procedure in the
master Lagrangian and take into account the ghost field to
preserve the BRS symmetry even after the gauge fixing, if
we follow the line of \cite{Kondo95th2}.  Therefore, in
order to show the quantum equivalence of the Thirring model
with some kind of gauge theory, we are required to find a
clever gauge-fixing so that the redundant fields can be
integrated out to arrive at the final gauge theory. Quite
recently, nevertheless, it was announced by Bralic et al.
\cite{BFMS95} that the non-Abelian version of the
(2+1)-dimensional Thirring model can be bosonized by
following the same strategy as that of Fradkin and
Schaposnik \cite{FS94} with help of the interpolating
Lagrangian of the form found in Karlhede et al.
\cite{KLRN87}.
The equivalent gauge theory obtained in \cite{BFMS95} is
somewhat similar to the Yang-Mills-Chern-Simons theory, but
does not exactly coincides with it.
However, the  questions raised above are not answered in
that paper, nor taken up are such questions.
In this sense, the bosonization of the (2+1)-dimensional
Thirring model is not yet well understood from our
viewpoint of gauge-invariant formulation.
\par
Finally we remark that, in the weak four-fermion coupling
limit $G \rightarrow 0$, the field $V_\mu$ decouples from
the master Lagrangian (\ref{master2}) after rescaling the
field $V_\mu$.  In this limit the master Lagrangian reduces
to
\begin{eqnarray}
 {\cal L}_M'[A_\mu,H_\mu] =
  {1 \over 2} \epsilon^{\mu_{1}...\mu_{D}}
H_{\mu_{3}...\mu_{D}} F_{\mu_{1}\mu_{2}}[A]
  + {\cal L}_G[A_\mu] .
\label{master4}
\end{eqnarray}
This should correspond to the free fermion model.
Actually this coincides with the result of \cite{BQ94}.
If we integrate out the $A_\mu$ field, we could
perform the bosonization of the free (!) fermion model
and would obtain the bosonized theory written in terms of
the field $H_\mu$.  However this simplified master
Lagrangian has the same problems as mentioned above in the
presence of the four-fermion interaction.  Therefore the
essential difficulty of non-Abelian bosonization comes not
only from the specific interaction of the original fermionic
model but also the gauge-invariance of
the master Lagrangian or the hidden gauge-invariance of the
free fermionic model.  The bosonization of the free
fermionic model was performed in the (1+1) dimensional case
by Burgess and Quevedo \cite{BQ94b}. However, it is not
straightforward to extend this analysis to the case of $D
>2$, since they use various peculiarities of (1+1)
dimensions.  In view of this, a detailed investigation
of the non-Abelian versions of the Thirring model will be
reported in a subsequent paper \cite{IKN95}.

\vskip 1cm
\section*{Acknowledgments}
K.-I. K. is supported in part by the Grant-in-Aid for
Scientific Research from the Ministry of Education, Science
and Culture (No.07640377).

\newpage
\section*{Figure Captions}
\begin{enumerate}
\item[Fig.1:]
 Equivalence of the various models.
\end{enumerate}

\newpage

\begin{center}
 \unitlength=1cm
  \begin{picture}(2.5,1)
   \thicklines
%%%           L[A,H,V]
  \put(-2,0){\framebox(8,1){  Master Lagrangian
      ${\cal L}[A,H,V]$\quad (\ref{master2})}}
%%%           L[A,H]
  \put(0,0){\vector(0,-1){3}}
  \put(-1.1,-2){$\int{\cal D}V$}
  \put(-4,-5){\framebox(5.4,2){\  ${\cal L}[A,H]
                \quad\qquad\qquad\qquad\qquad $ }}
  \put(-2.2,-3.6){$D=2\ (\ref{eq:kondo2d})$}
  \put(-2.2,-4.1){$D=3\ (\ref{3LinterK})\
                (\ref{3LinterK-next})$}
  \put(-2.2,-4.6){$D=4\ (\ref{4DL[A,H]})$}
%%%           L[H,V]
  \put(3,0){\vector(0,-1){3}}
  \put(3.1,-2){$\int{\cal D}A$}
  \put(2,-5){\framebox(6,2){\  ${\cal L}[H,V]
         \qquad \qquad\qquad\qquad\qquad $ }}
  \put(4.0,-3.7){$D=3\ (\ref{interFS})  \
                (\ref{3DinterFSnext})$}
  \put(4.0,-4.3){$D=4\ (\ref{4DL[H,V]})$}
%%% non-linear sigma
  \put(-2,0.5){\line(-1,0){3}}
  \put(-5,0.5){\vector(0,-1){10.5}}
  \put(-6,-12){\framebox(4,2){}}
  \put(-5.7,-10.5){Gauged Non-Linear}
  \put(-5,-11.1){$\sigma$ Model\ (\ref{NLSM})}
  \put(-4.5,-11.6){${\cal L}[A,\varphi]$}
  \put(-4.9,-0.5){$\int{\cal D}V\int{\cal D}H$}
%%% Thirring model
  \put(-5,-14.5){\vector(0,1){2.5}}
  \put(-6.7,-13.3){$\int{\cal D}\psi{\cal D}\ol{\psi}$}
  \put(1,-14.5){\vector(0,1){2.5}}
  \put(1.1,-12.8){$\int{\cal D}\psi{\cal D}\ol{\psi}$}
  \put(1.1,-13.8){Auxiliary Field $A_{\mu}$}
  \put(-6,-16){\framebox(10,1.5){}}
  \put(-3.3,-15){Massive Thirring Model}
  \put(-2.3,-15.6){${\cal L}[\psi,\ol{\psi}]$\quad
(\ref{th})}
  \put(-4.8,-13.3){Auxiliary Field $A_{\mu}$}
  \put(-4.8,-13.8){St\"uckelberg Field $\theta$}
%%% L[H]
  \put(0.5,-5){\vector(0,-1){2}}
  \put(0.6,-6){$\int{\cal D}A$}
  \put(2.5,-5){\vector(0,-1){2}}
  \put(2.6,-6){$\int{\cal D}V$}
  \put(0,-9.5){\framebox(7,2.5){ }}
  \put(1,-7.6){Bosonized Theory\quad ${\cal L}[H]$}
  \put(1.3,-8.1){$D=2\ $(\ref{scalar})}
  \put(1.3,-8.6){$D=3\ $(\ref{MCS})}
  \put(1.3,-9.1){$D=4\ $(\ref{L[H]2})\ }
%%% L[A]
  \put(-0.5,-5){\vector(0,-1){5}}
  \put(-1.6,-6){$\int{\cal D}H$}
  \put(-1.5,-12){\framebox(5,2){\quad ${\cal L}[A]
      \qquad\qquad\qquad\qquad\qquad $}}
  \put(0.2,-10.6){$D=2\ $(\ref{2dL[A]})}
  \put(0.2,-11.1){$D=3\ $(\ref{selfdual})}
  \put(0.2,-11.6){$D=4\ $(\ref{L[A]})}
 \end{picture}
\end{center}

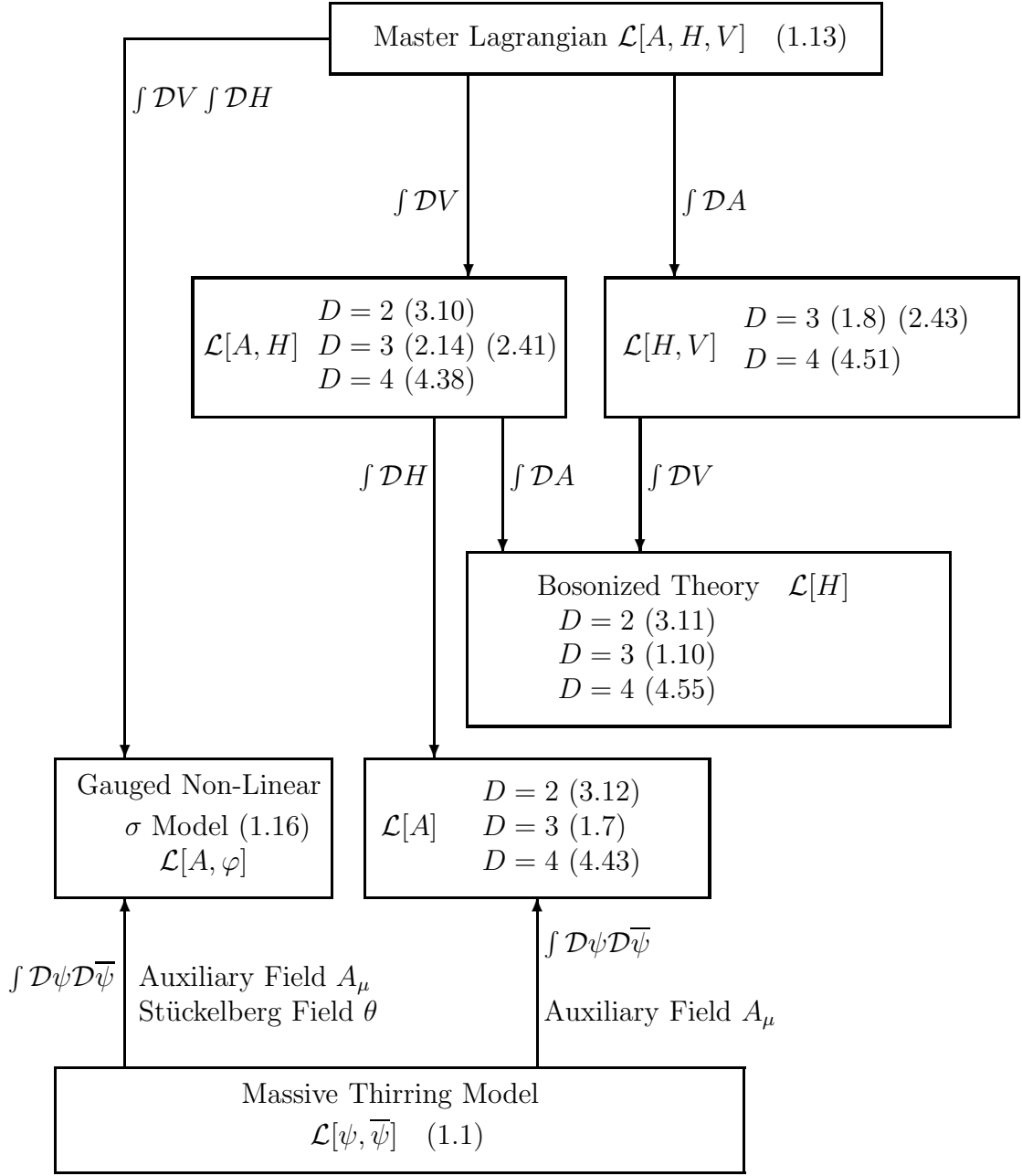
\end{document}